\documentstyle[prc,aps,preprint,psfig]{revtex}
\draft
\begin{document}

\title{Baryon stopping and strange baryon/antibaryon production at SPS
energies}
\author{H. Weber, E. L. Bratkovskaya\thanks{Supported by DFG}
 and H. St\"ocker\\[2mm]
{\normalsize Institut f\"{u}r Theoretische Physik,
Universit\"{a}t Frankfurt}\\
{\normalsize 60054 Frankfurt am Mine, Germany}}
%\date{ }
\maketitle

\begin{abstract}
The amount of proton stopping in central $Pb+Pb$ collisions from 20 --
160 A$\cdot$GeV as well as hyperon and antihyperon rapidity distributions are
calculated within the UrQMD model in comparison to experimental data at
40, 80 and 160 A$\cdot$GeV taken recently from the NA49 collaboration.
Furthermore,  the amount of baryon stopping at 160 A$\cdot$GeV for
$Pb+Pb$ collisions is studied as a function of centrality in comparison
to the NA49 data.  We find  that the strange baryon yield is reasonably described
for central collisions, however, the rapidity
distributions are somewhat more narrow than the data.  Moreover, the
experimental antihyperon rapidity distributions at 40, 80 and 160
A$\cdot$GeV are underestimated by up to factors of 3 - depending on the annihilation
cross section employed -  which might be
addressed to missing multi-meson fusion channels in the UrQMD model.
\end{abstract}

\vspace{0.3cm}\noindent
PACS{{ 25.75.+r} { Relativistic heavy-ion collisions} }

%---------------------------------------------------------------------
\newpage
\narrowtext
%\twocolumn
\section{Introduction}

Present lattice QCD calculations indicate that strongly interacting
hadronic matter at temperatures of 150-170 MeV (or energy densities of
1-2 GeV/fm$^3$) should undergo a phase transition to a new state of
matter generally denoted as quark-gluon-plasma (QGP). It is also common
believe that this state of matter existed during the early phase of the
universe until the temperature drop due to the rapid expansion lead to
the freeze-out of hadrons which constitute a sizeable fraction of the
total mass of the universe. Whereas the 'big bang' has only been a
single event - for presently living observers - relativistic collisions
of heavy nuclei offer the unique possibility to study the dynamics of a
huge number of 'tiny bangs' under well controlled laboratory
conditions. Hadronic spectra and relative hadron abundancies reflect
the dynamics in the hot and dense zone formed in the early phase of the
reaction.

Whereas meson rapidity distributions and transverse mass spectra
essentially reflect the dynamics of newly produced $q\bar{q}$ pairs,
the baryon rapidity and transverse mass distributions give important
information on baryon stopping \cite{Scheid} whereas antibaryon
abundancies shed some light on quark chemical potentials $\mu_q$ at the
space-time points of chemical decoupling, i.e. when chemical reactions
no longer occur due to a large average separation between the hadrons.
The latter statement, however, only holds if an approximate chemical
equilibrium is reached in the collision zone of nucleus-nucleus
reactions. In fact, chemical equilibrium models -- based on
extrapolations of existing data at the AGS and SPS -- suggest that the
highest strange baryon abundancies should occur in central collisions
of heavy nuclei between 20 and 40 A$\cdot$GeV \cite{StatMod}.
Furthermore, the degree of baryon stopping is related (by energy-momentum conservation)
to the number of newly produced hadrons $dN/dy$ (per unit rapidity) which can be used to extrapolate
 the achieved energy density in these collisions by adopting the
Bjoken formula \cite{bjorken}
\begin{equation}
\label{bjor}
\epsilon = \frac{M_T}{\tau_0 A} \left.\frac{dN}{dy}\right|_{y=y_{cm}},
\end{equation}
where $A$ is the transverse (geometrical) overlap region, $M_T$ is the
average transverse mass and $\tau_0$ the proper production time which
is estimated to be in the order of 1 fm/c. According to (\ref{bjor})
the energy densities reached in central $Pb+Pb$ collisions at the SPS
-- using experimental information on $M_T$ and ${dN(y_{cm})}/{dy}$
should be in the order of 2.5--3.5 GeV/fm$^3$, i.e. well above the
critical energy density for a transition to a QGP in equilibrium.

The data from the SPS on baryon stopping demonstrate that simple
extrapolations from $pp$ collisions at the same energy do not show
enough baryon stopping (cf. e.g. \cite{QM98,QM00}). Here transport
models employing hadronic and string degrees-of-freedom such as RQMD
\cite{RQMD}, UrQMD \cite{UrQMD1,UrQMD2} or HSD \cite{HSD_K,CBRep98} do
a better job since  the formation and multiple rescattering of formed
hadrons are included in these approaches. Furthermore, such transport
calculations allow to study the change in dynamics from elementary
baryon-baryon or meson-baryon collisions to proton-nucleus reactions
or from peripheral to central nucleus-nucleus collisions in a unique way
without changing any parameter. This is of central importance since the
prejudice of thermal and chemical equilibrium does not hold in all of
these reactions and the transport studies allow to explore the amount
of (thermal or chemical) equilibrium reached in such collisions
\cite{Brat99,Bravina}.

Experimentally, the dynamics of heavy nucleus-nucleus collisions have
been studied up to 11.6 A$\cdot$GeV at the BNL AGS and an extensive
program has been carried out at the 'top' CERN SPS energy of 160
A$\cdot$GeV, whereas the intermediate range from 11 to 160
A$\cdot$GeV has been practically unexplored from the experimental side.
Only recently, experiments for $Pb + Pb$ collisions
at 40 and 80 A$\cdot$GeV have been performed at the CERN SPS
\cite{NA49_new,NA49_Lam} and further experimental
measurements are expected at 20 A$\cdot$GeV \cite{SPS20}.
In this respect there is considerable hope that
the experimental data can throw light
on the basic question -- if we might find
signatures for an intermediate QGP state or if we just see strongly
interacting hadronic matter.

In a previous study -- within the UrQMD approach -- we have addressed
pion, kaon and antikaon abundancies and spectra in central $Pb+Pb$
collisions from 20 -- 160 A$\cdot$GeV in comparison to the  data from
the NA49 Collaboration \cite{Weber_K}. In general, we have found that
the UrQMD model reasonably describes the data, however, systematically
overpedicts the $\pi^-$ yield by $\sim 20$\%, whereas the $K^+$ yield
is underestimated by $\sim 15$\%. The $K^-$ yields are in a good
agreement with the data for all energies.  This suggests that the
production of antistrange quarks ($\bar{s}$) might be somewhat low in
the transport model (as in the HSD approach \cite{HSD_K}) whereas the
production of the lightest $q\bar{q}$ pairs is overestimated
systematically. However, in order to obtain a complete information on
the abundancy of $s,\bar{s}$ quarks one has to study strange baryon
production and antihyperon production, too, since strangeness
conservation implies the same amount of $s$ and $\bar{s}$ quarks to be
produced in the collision. It is the aim of this work to provide an
answer to this question within nonequilibrium transport theory.

%-------------------------------------------------------------
\section{Proton stopping and hyperon production}

The UrQMD transport approach is described in Refs. \cite{UrQMD1,UrQMD2}
and includes all baryonic resonances up to an invariant mass of 2 GeV
as well as mesonic resonances up to 1.9 GeV as tabulated in the PDG
\cite{PDG}. For hadronic continuum excitations we employ a string model
with meson formation times in the order of 1-2~fm/c depending on the
momentum and energy of the created hadrons.  The transport approach is
matched to reproduce the nucleon-nucleon, meson-nucleon and meson-meson
cross section data in a wide kinematical regime \cite{UrQMD1,UrQMD2}.
At the high energies considered here the particles are essentially
produced in primary high energetic collisions by string excitation and
decay, however, the secondary interactions among produced particles
(e.g.  pions, nucleons and excited baryonic and mesonic resonances)
also contribute to the particle dynamics -- in  production as well as
in absorption.

\begin{figure}[h]
\centerline{\psfig{figure=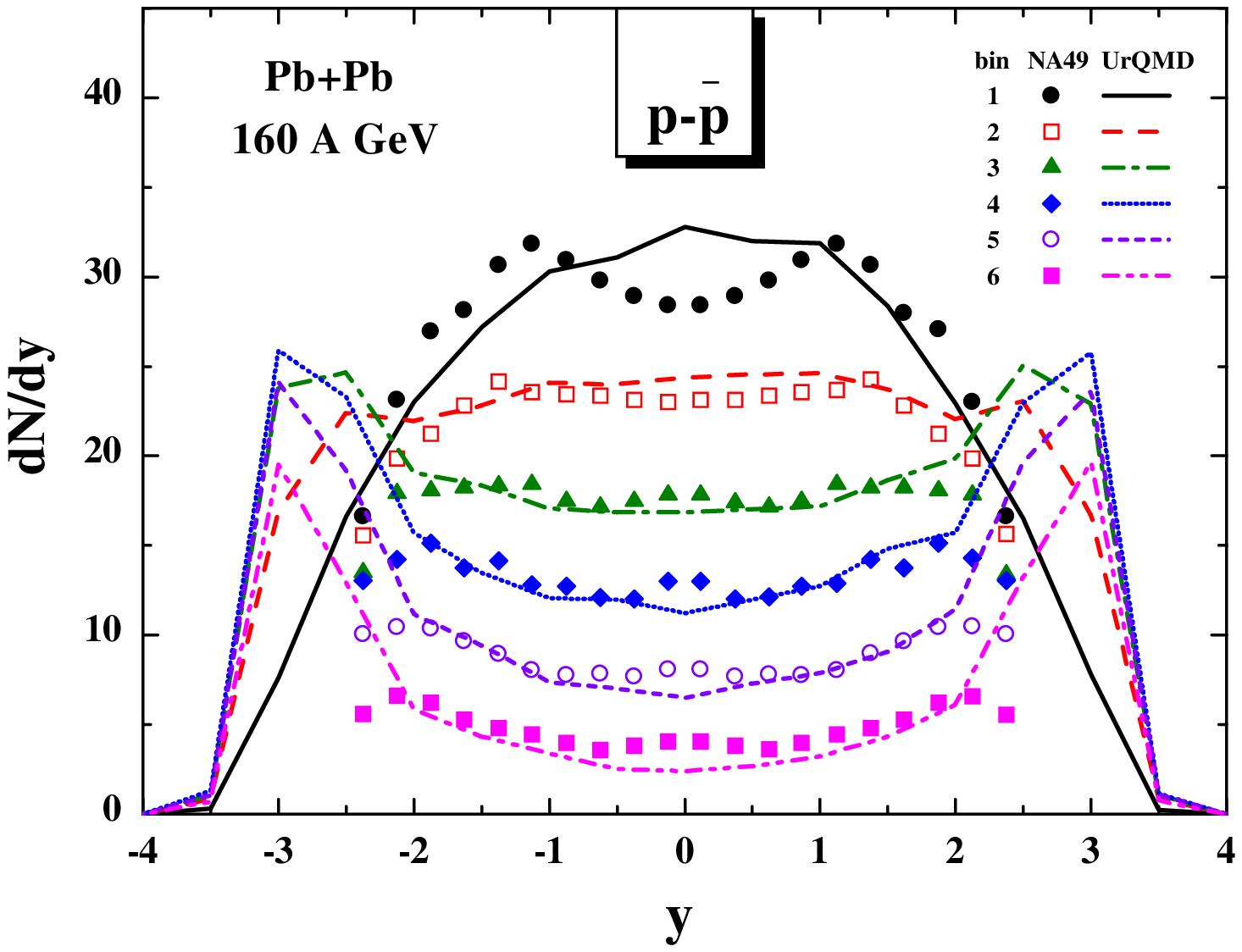,width=15cm}}
\vspace*{5mm}
\caption{The rapidity distribution of net protons $p-\bar p$
in $Pb + Pb$ collisions at 160 A$\cdot$GeV calculated
within the UrQMD model (lines) in comparison to the experimental
data from the NA49 Collaboration  \protect\cite{NA49pNew} for 6
different centrality classes -- from the most
central (bin 1) to the very peripheral collisions (bin 6).}
\label{prot160}
\end{figure}

Here we can come directly to the results for baryons and antibaryons
and start at the highest bombarding energy of 160 A$\cdot$GeV. The
comparison of the UrQMD results on baryon stopping for the most central
$Pb + Pb$ collisions at 160 A$\cdot$GeV to the NA49 data
\cite{NA49pOld} has been reported previously in Refs.
\cite{UrQMD1,Bass99}. In Fig. \ref{prot160} we compare the UrQMD
(version 1.3) calculations for the net proton rapidity distribution
$p-\bar{p}$ to the most resent data from the NA49 Collaboration
\cite{NA49pNew} for 6 different centrality classes of $Pb+Pb$
collisions -- from the most central (bin 1) to the very peripheral
collisions (bin 6).  Note, that the spectators are excluded from the
calculated $dN/dy$ spectra in line with the experimental measurement.
We find that the UrQMD model overestimates the stopping for the most
central rapidity bin, i.e. the data show a slight dip at midrapidity
and a two peak stucture, which indicates that  full stopping is not
achieved at 160 A$\cdot$GeV even for this heavy system.  On the other
hand, it is quite remarkable that the hadron/string approach well
reproduces the $p-\bar{p}$ rapidity distributions as a function of
centrality.

\begin{figure}[ht]
\centerline{\psfig{figure=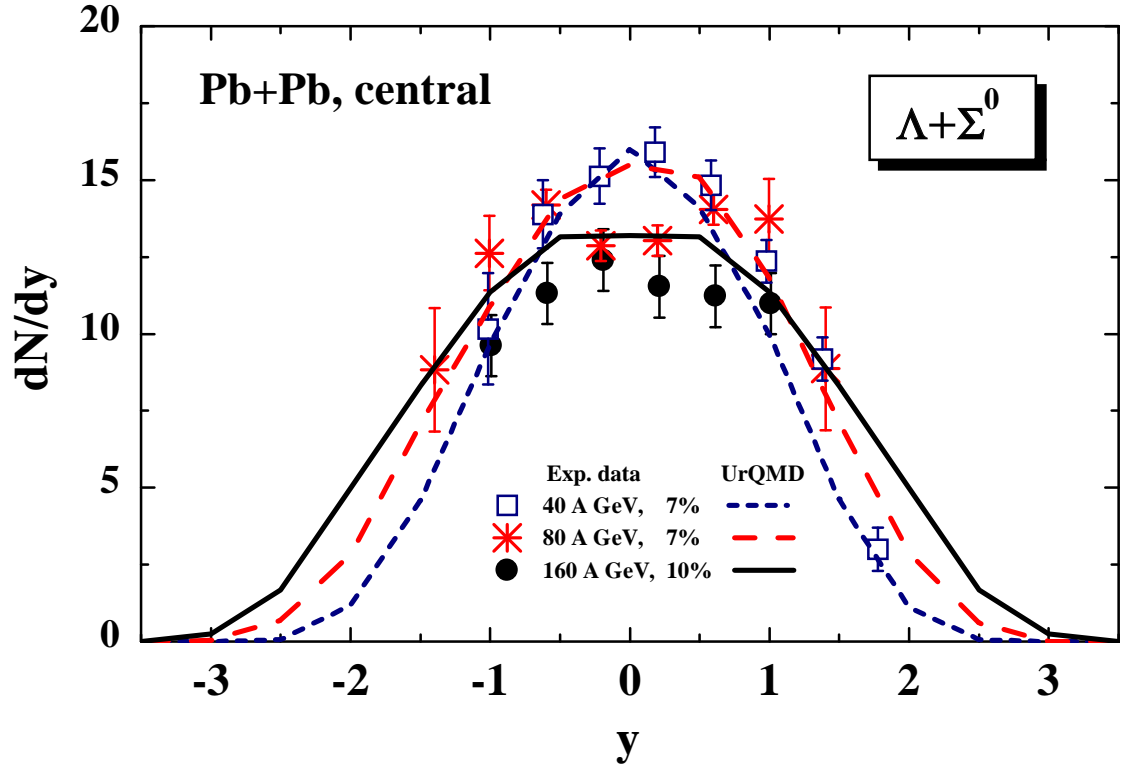}}
\vspace*{5mm}
\caption{The UrQMD calculations of the hyperon ($\Lambda + \Sigma^0$)
rapidity distributions for $Pb + Pb$ collisions at 40 (7\% central), 80
(7\% central) and 160 (10\% central) A$\cdot$GeV in comparison to the
data from the NA49 Collaboration \protect\cite{NA49_Lam}.}
\label{lamsps}
\end{figure}

We step on with the hyperon ($\Lambda + \Sigma^0$) rapidity
distributions at 40, 80 and 160 A$\cdot$GeV in comparison to the data
from NA49 \cite{NA49_Lam} -- Fig. \ref{lamsps}. The UrQMD calculations
show an increasing hyperon yield with bombarding energy essentially due
to a broadening of the rapidity distribution, while the midrapidity
distributions at 40 and 80 A$\cdot$GeV are practically the same.  The
data from the NA49 Collaboration show a decreasing hyperon yield at
midrapidity with higher bombarding energy while suggesting a slightly
larger width in $dN/dy$. Note, however, that the data at 160
A$\cdot$GeV correspond to 10\% centrality whereas the lower energies
are for 7\% centrality, respectively. We mention that for 7\%
centrality our calculations at 160 A$\cdot$GeV roughly give the same
$\Lambda + \Sigma^0$ yield at midrapidity than for the lower energies
of 40 and 80 A$\cdot$GeV. It is not clear at present from the data, if
the total integrated yields are compatible with our calculations.
However, as demonstrated in Ref.  \cite{Weber_K}, the UrQMD model
describes rather well the antikaon rapidity distributions from 40 --
160 A$\cdot$GeV whereas the kaon rapidity distributions are
underestimated by about 15\%.  Consequently, by strangeness
conservation, which is strictly fulfilled in the UrQMD approach, the
hyperon yield should also be underestimated slightly.

\begin{figure}[ht]
\centerline{\psfig{figure=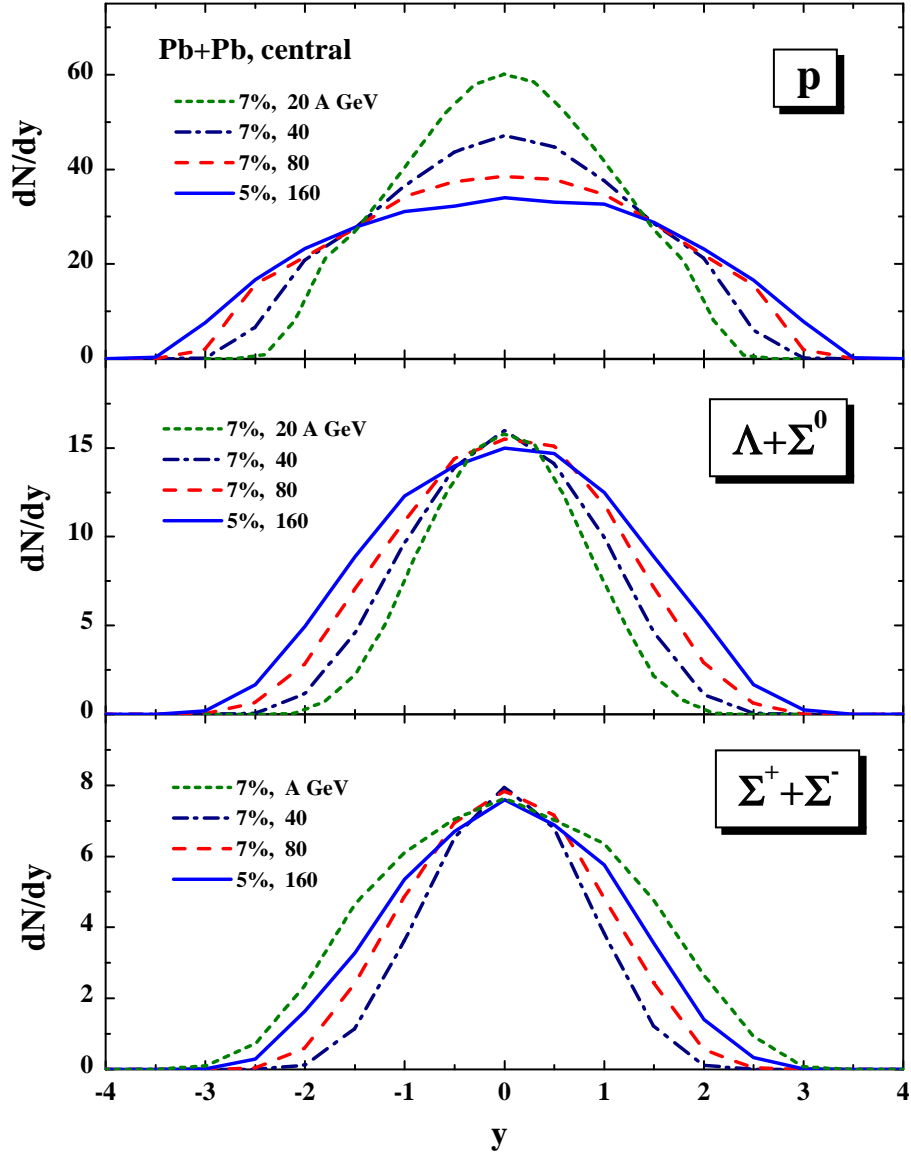,width=12cm}}
\vspace*{5mm}
\caption{The rapidity distributions of net protons $p-\bar p$, hyperons
($\Lambda + \Sigma^0$  and $\Sigma^+ + \Sigma^-$) calculated within the
UrQMD model for  7\% central $Pb + Pb$ collisions at 20 (short dashed
lines), 40 (dot-dashed lines), 80 (dashed lines) and for 5\% central collisions at
160 A$\cdot$GeV (solid lines).}
\label{p_bar}
\end{figure}

We, furthermore, provide an overview on  rapidity distributions of
protons, neutral ($\Lambda + \Sigma^0$) and charged hyperons ($\Sigma^+
+\Sigma^-$) at 40, 80 and 160 A$\cdot$GeV from 7\% or 5\% central
$Pb+Pb$ collisions within the UrQMD model as well as predictions for 20
A$\cdot$GeV (Fig. \ref{p_bar}), where experimental measurements will be
taken in near future \cite{SPS20}. Whereas the net proton density at
midrapidity decreases strongly with higher bombarding energy -- which
should be attributed to a lower amount of baryon stopping -- the width
in rapidity increases accordingly since the net $p-\bar{p}$ number is a
constant, if the produced meson system on average is charge neutral.
The situation with strange baryons is different since a newly produced
$s$-quark is contained in their wave function. In the UrQMD transport
model this leads to a much narrower rapidity distribution for strange
baryons than for protons from 20 -- 160 A$\cdot$GeV as seen from Fig.
\ref{p_bar}. Consequently, the $\Lambda/p$ ratio varies sensitively
with rapidity.

\section{Antiproton and antihyperon production}

\begin{figure}[ht]
\centerline{\psfig{figure=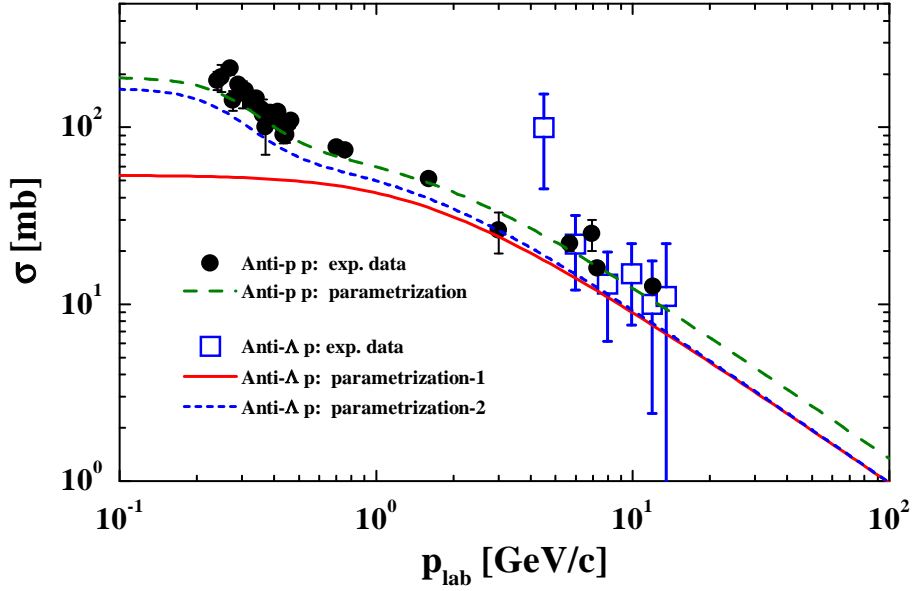,width=12cm}}
\vspace*{5mm}
\caption{The annihilation cross section of $\bar p$ and $\bar \Lambda$
with nucleons as a function of incident ($\bar p$ or $\bar\Lambda$)
momentum in the laboratory frame. The solid circles are the $\bar p$ data
from \protect\cite{PDG}, the open squares correspond to the
$\bar \Lambda p$ data from \protect\cite{Eisele}.
The dashed line is the parametrization of the $\bar p$
annihilation cross section used in UrQMD, the short dashed and solid lines
correspond to the two different parametrizations of the $\bar\Lambda p$
data (see text).}
\label{antil-xs}
\end{figure}

We continue with antibaryon production in central $Pb+Pb$ collisions at
SPS energies. Since the final $\bar{p}$ or $\bar{\Lambda}$ rapidity
distributions are sensitive to their annihilation cross section with
nucleons, we first discuss the actual implementation of annihilation
within UrQMD.  In this respect we show in Fig.
\ref{antil-xs} the annihilation cross section of $\bar p$ and $\bar
\Lambda$ with nucleons as a function of the incident ($\bar p$ or
$\bar\Lambda$) momentum in the laboratory frame. The solid circles are
the $\bar p$ data from \protect\cite{PDG} while the open squares
correspond to the $\bar \Lambda p$ data from \cite{Eisele}.  The dashed
line stands for the parametrization of the $\bar p$ annihilation cross
section used in UrQMD while the short dashed and solid lines correspond
to  two different parametrizations of the $\bar\Lambda p$ annihilation
cross section, which are both compatible to the experimental data (open
squares), however, involve quite different extrapolations to the low
momentum regime. The parametrization-1 (short dashed line) assumes
\begin{equation}
\sigma^{ann}_{\bar{\Lambda} N} (\sqrt{s})
	\approx 0.8 \ \sigma^{ann}_{\bar{p} N} (\sqrt{s}),
\label{param1}
\end{equation}
thus relating the different cross sections at the same invariant
energy $\sqrt{s}$, which leads to a constant annihilation cross
section for antilambdas at low momentum of $\approx$ 55 mb
(default in UrQMD). The parametrization-2 (solid line) instead assumes
\begin{equation}
\sigma^{ann}_{\bar{\Lambda} N} (p_{lab})
	\approx 0.8 \ \sigma^{ann}_{\bar{p} N} (p_{lab}),
\label{param2}
\end{equation}
thus relating the different cross sections at the same laboratory
momentum $p_{lab}$. We note again that the data on $\bar{\Lambda}$
annihilation at high momenta are compatible with both parametrizations.

\begin{figure}[ht]
\centerline{\psfig{figure=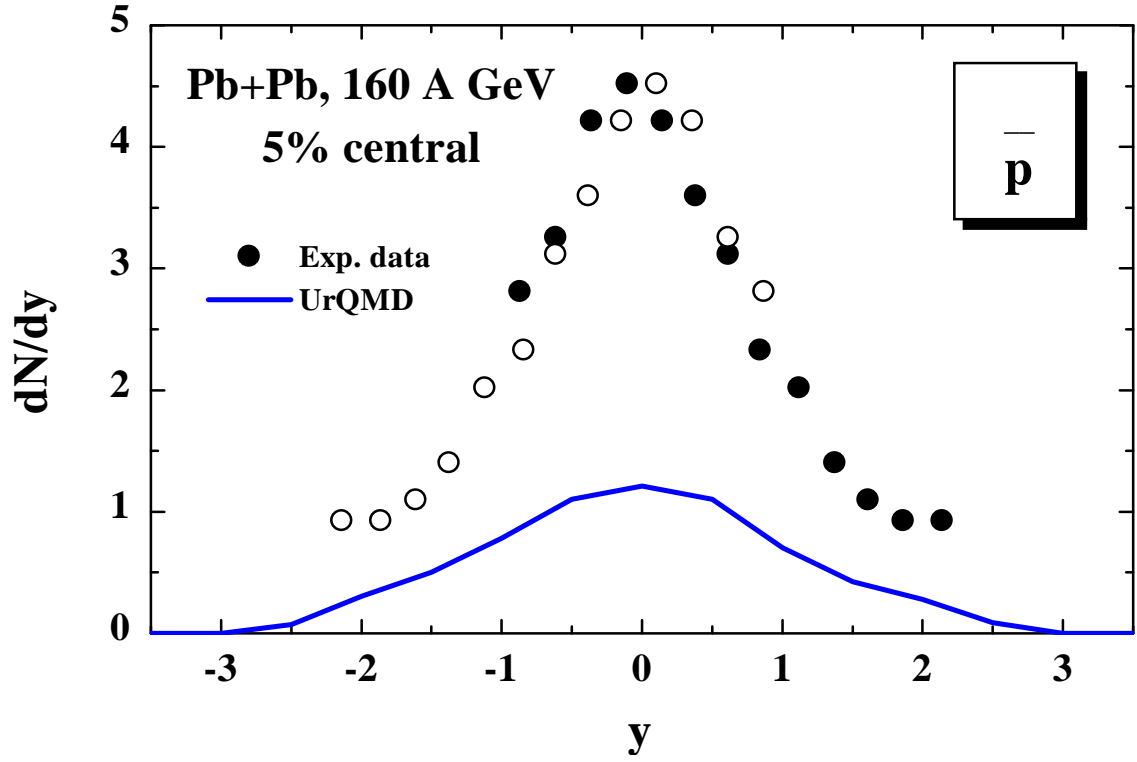,width=15cm}}
\vspace*{5mm}
\caption{The UrQMD calculations of the antiproton ($\bar p$)
rapidity distribution for 5\% central $Pb + Pb$
collisions at 160 A$\cdot$GeV in comparison to the data from the
NA49 Collaboration \protect\cite{NA49_pbar}.}
\label{pbar160}
\end{figure}

The UrQMD calculations of the antiproton ($\bar p$) rapidity
distribution for 5\% central $Pb + Pb$ collisions at 160 A$\cdot$GeV
are shown in Fig. \ref{pbar160} in comparison to the data from the NA49
Collaboration \cite{NA49_pbar}, which also include some contribution
from the feeddown of $\bar{\Lambda}$ and $\bar{\Sigma}^0$. The
experimental distribution is underestimated severely in UrQMD
suggesting either a much lower annihilation cross section for
antiprotons or the dominance of multi-meson fusion channels as
suggested in Refs. \cite{Rapp,carsten,wolfgang}.

\begin{figure}[ht]
\centerline{\psfig{figure=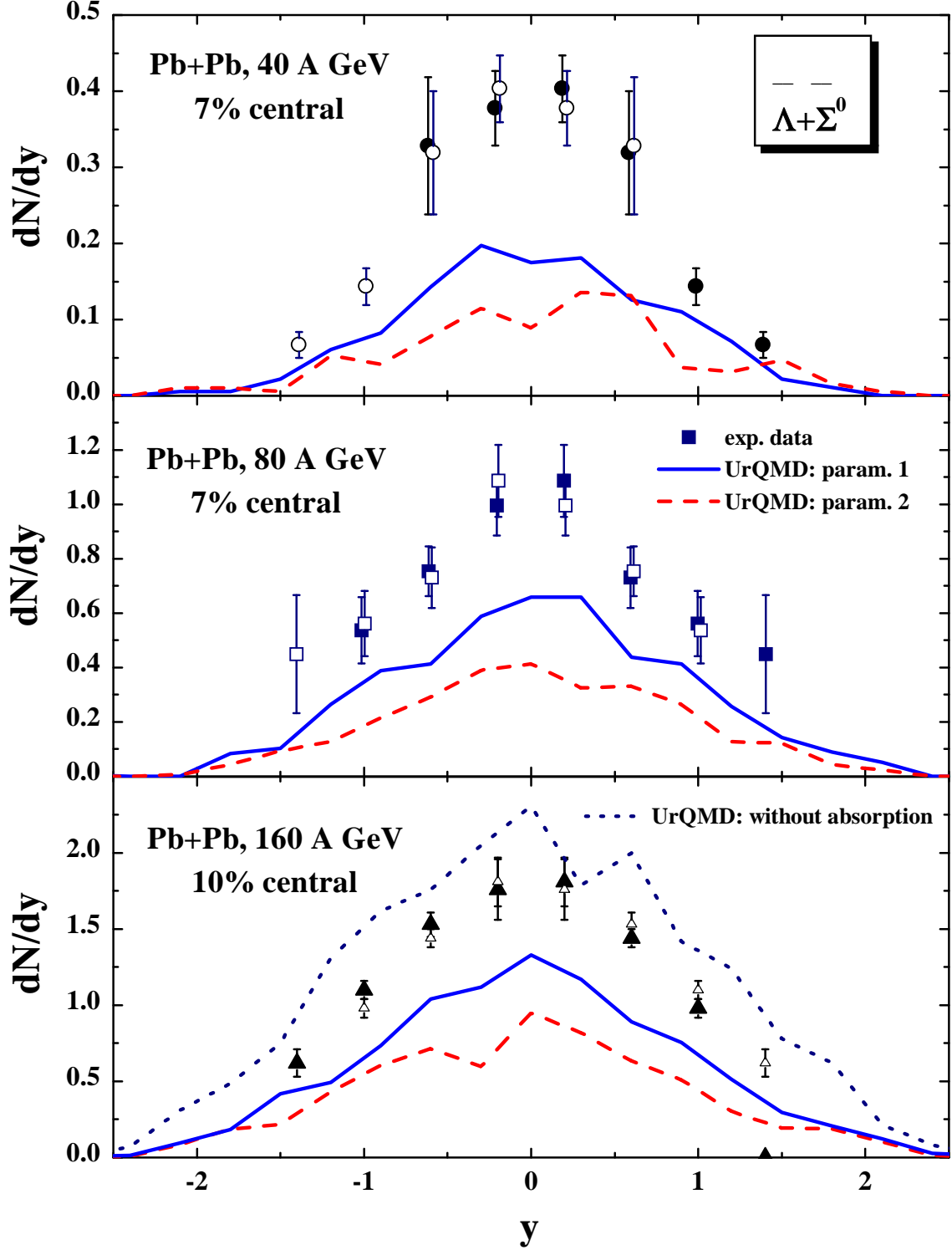,width=15cm}}
\vspace*{5mm}
\caption{The UrQMD calculations of the antihyperon ($\bar\Lambda +
\bar\Sigma^0$) rapidity distributions for 7\% central $Pb + Pb$
collisions at 40 and 80 A$\cdot$GeV and for 10\% central $Pb + Pb$
at 160 A$\cdot$GeV in comparison to the data from the NA49 Collaboration
\protect\cite{NA49_Lam,NA49_QM02}. The short-dashed line for 160 A$\cdot$GeV
corresponds to a calculation without antihyperon annihilation.}
\label{alamsps}
\end{figure}

Within the strangeness balance discussed in the context with Figs. 2
and 3, the antistrangeness content of antihyperons ($\bar{\Lambda} +
\bar{\Sigma}$) has been neglected. This conjecture remains to be
proven. In fact, as shown in Fig. \ref{alamsps}, the experimental data
\cite{NA49_Lam,NA49_QM02} for central $Pb+Pb$ collisions at 80
A$\cdot$GeV give $dN/dy \approx$ 1, which is within the experimental
error bars for $\Lambda+\Sigma^0$ in Fig. \ref{lamsps}.  This even more
holds true at the lower bombarding energy of 40 A$\cdot$GeV.  The UrQMD
calculations for the same centrality bin  underestimate the NA49 data
\cite{NA49_Lam} by about of a factor of 2 (for parametrization-1) or 3
(for parametrization-3) at 40 and 80 A$\cdot$GeV whereas the data at
160 A$\cdot$GeV are underestimated only by about factors of 1.5-2. The
short-dashed line (in the lower part for 160 A$\cdot$GeV) shows the
result of a calculation without antihyperon annihilation which only
slightly overestimates the data. When integrating over rapidity we find
that in case of parametrization-1 about half of the antihyperons are
annihilated whereas for the parametrizarion-2  $\sim$2/3 of the
antihyperons disappear.

It has been shown previously in Ref.  \cite{Soff99} that the standard
UrQMD model (with parametrization-1) underestimates the (multi-)strange
baryon multiplicity for central $Pb + Pb$ at 160 A$\cdot$GeV. As argued
in \cite{Soff99}, the inclusion of non-hadronic medium effects, like
color-ropes \cite{colroper} (simulated in UrQMD by increasing the
string-tension), enhances the multiplicity of (anti-)strange baryons.
The missing antihyperon yield can be attributed also to multi-meson
fusion channels involving $K, \bar{K}, K^+, \bar{K}^*$ mesons
\cite{carsten,wolfgang} that are not accounted for in the calculations
reported here. Furthermore, the high abundance of $\Omega$ and
$\bar{\Omega}$ seen experimentally might also signal the appearance of
disoriented chiral condensates (DCC's) as put forward by Kapusta and
Wong \cite{kapusta}. In short, this issue is presently still open.

\begin{figure}[ht]
\centerline{\psfig{figure=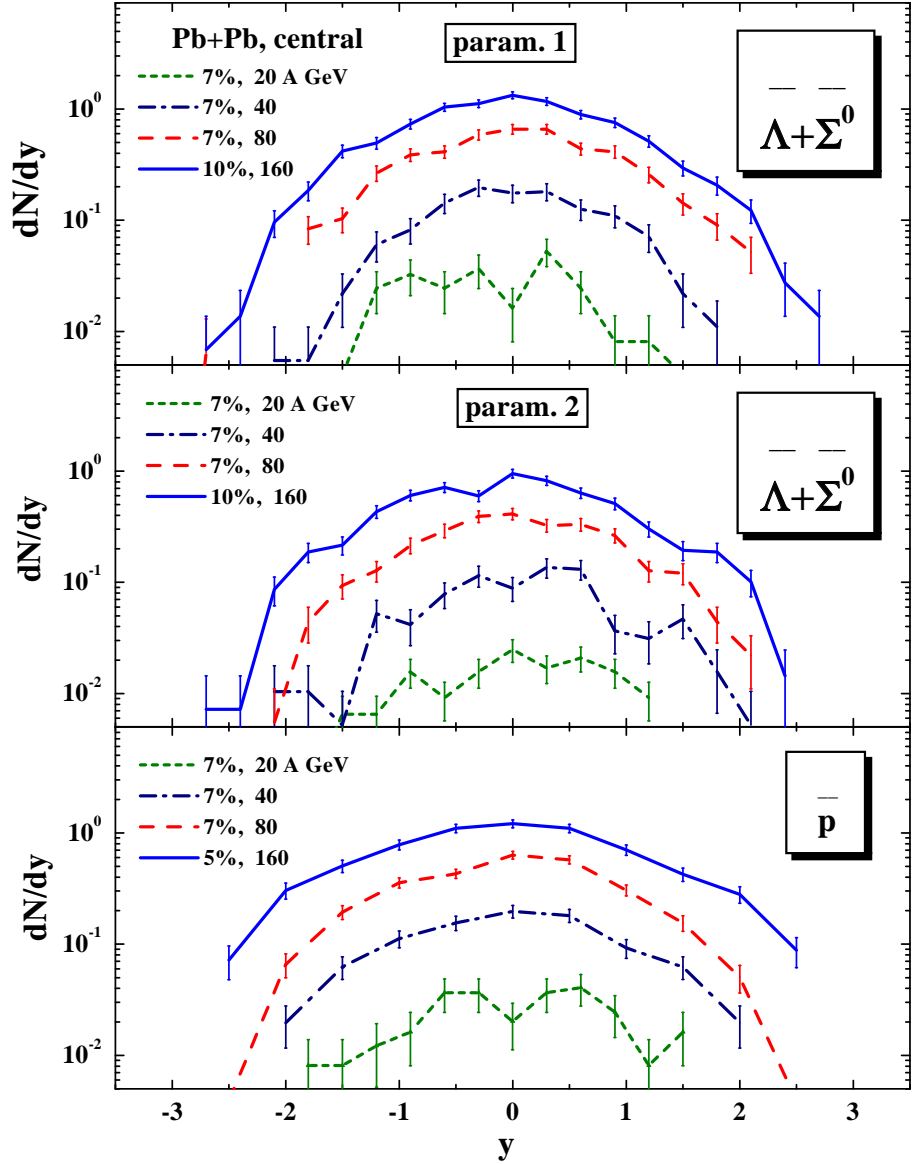,width=12cm}}
\vspace*{5mm}
\caption{The rapidity distributions of antihyperons ($\bar\Lambda +
\bar\Sigma^0$ ) and antiprotons ($\bar p$) calculated within the UrQMD
model for 7\% central $Pb + Pb$ collisions at 20 (short dashed lines),
40 (dot-dashed lines), 80 (dashed lines) and for 10\% central ($\bar\Lambda + \bar\Sigma^0$)
  or 5\% central collisions (for $\bar p$) at 160 A$\cdot$GeV
(solid lines). The upper plot corresponds to the 'parametrization-1'
for $\bar \Lambda p$ annihilation cross section whereas the middle
plot shows the UrQMD results with the 'parametrization-2' (see text).}
\label{y-anti}
\end{figure}

In order to provide an overview on antiproton and antihyperon
production (in analogy to Fig. 3) we show in Fig. \ref{y-anti} the
 rapidity distributions of antihyperons ($\bar\Lambda + \bar\Sigma^0$)
calculated within the UrQMD model for 7\% central $Pb + Pb$ collisions
at 20 (short dashed lines), 40 (dot-dashed lines), 80 (dashed lines)
and for 10\% central ( $\bar\Lambda + \bar\Sigma^0$)  and 5\% central
($\bar p$) collisions at 160 A$\cdot$GeV (solid lines). The upper plot
corresponds to the 'parametrization-1' for $\bar \Lambda p$
annihilation cross section whereas the middle plot shows the UrQMD
results with the 'parametrization-2'.  The abundancy of strange
antibaryons ($\bar{\Lambda} + \bar{\Sigma}^0$) increases rapidly with
bombarding energy. Note, since the antihyperon yield is very low
especially at 20 A$\cdot$GeV, we present the antihyperon rapidity
distribution in a logarithmic scale
and indicate the statistical errorbars in order to demonstrate the
accuracy/statistics achieved in the UrQMD calculations. As discussed
above the antihyperon absorption is more pronounced for
parametrization-2 especially at lower bombarding energy.  We note in
passing, taht the (rapidity integrated) $\bar{\Lambda}/\bar{p}$ ratio
from the UrQMD calculation is $\sim$ 0.9 and 0.6 for all bombarding energies
from 20 - 160 A$\cdot$GeV within the parameterset-1 and 2, respectively.

%-------------------------------------------------------------
\section{Conclusions}

In summary, we have calculated the amount of baryon stopping in central
$Pb+Pb$ collisions from 20 -- 160 A$\cdot$GeV as well as hyperon rapidity
distributions in comparison to experimental data at 40, 80 and 160
A$\cdot$GeV taken recently by the NA49 collaboration \cite{NA49_Lam}.
We have demonstrated, furthermore, that the UrQMD model reasonably
reproduces the amount of baryon stopping at 160 A$\cdot$GeV for $Pb+Pb$
collisions as a function of centrality. The comparison of our
calculations  for hyperons with the experimental data, however,
indicates that the strange baryon yield at midrapidity is slightly
overestimated whereas the calculated rapidity distributions are
somewhat more narrow than the data. This discrepancy might indicate a
different mechanism for strange hyperon production than the string
mechanism in the transport model. On the other hand, the experimental
antihyperon rapidity distributions at 40, 80 and 160 A$\cdot$GeV as well as
the antiproton rapidity distribution at 160 A$\cdot$GeV are
underestimated by up to factors of 3 which we address to missing
multi-meson fusion channels \cite{Rapp,carsten,wolfgang} in the UrQMD model.
Note, however, that instead of multi-meson fusion channels also
disoriented chiral condensates might explain the enhanced
production of multistrange baryons as suggested in Ref.
\cite{kapusta}.

%------------------------------------------------------------------------

%--------------------------------------------------------------

\clearpage

\clearpage

\clearpage

\clearpage

\clearpage

\clearpage

\end{document}